\documentclass{aastex61}

\newcommand{\adb}{\nabla_{\rm ad}}
\newcommand{\hpeb}{H_{\rm peb}}
\newcommand{\hgas}{H_{\rm gas}}
\newcommand{\kdust}{\kappa_{\rm dust}}
\newcommand{\mtot}{M_{\rm tot}}
\newcommand{\st}{\rm St}
\newcommand{\stcrit}{{\rm St}_{\rm crit}}
\newcommand{\tstop}{t_{\rm stop}}
\newcommand{\vrel}{v_{\rm rel}}

\received{}
\revised{}
\accepted{\today}
\submitjournal{ApJ}

\shorttitle{Steamworlds}
\shortauthors{Chambers}

\begin{document}

\title{Steamworlds: Atmospheric Structure and Critical Mass of Planets Accreting Icy Pebbles}

\correspondingauthor{John Chambers}
\email{jchambers@carnegiescience.edu}

\author{John Chambers}
\affil{Carnegie Institution for Science \\
Department of Terrestrial Magnetism, \\
5241 Broad Branch Road, NW, \\
Washington, DC 20015, USA}

\begin{abstract}
In the core accretion model, gas-giant planets first form a solid core, which then accretes gas from a protoplanetary disk when the core exceeds a critical mass. Here, we model the atmosphere of a core that grows by accreting ice-rich pebbles. The ice fraction of pebbles evaporates in warm regions of the atmosphere, saturating it with water vapor. Excess water precipitates to lower altitudes. Beneath an outer radiative region, the atmosphere is convective, following a moist adiabat in saturated regions due to water condensation and precipitation. Atmospheric mass, density and temperature increase with core mass. For nominal model parameters, planets with core masses (ice + rock) between 0.08 and 0.16 Earth masses have surface temperatures between 273 K and 647 K and form an ocean. In more massive planets, water exists as a super-critical convecting fluid mixed with gas from the disk. Typically, the core mass reaches a maximum (the critical mass) as a function of the total mass when the core is 2-5 Earth masses. The critical mass depends in a complicated way on pebble size, mass flux, and dust opacity due to the occasional appearance of multiple core-mass maxima. The core mass for an atmosphere of 50 percent hydrogen and helium may be a more robust indicator of the onset of gas accretion. This mass is typically 1-3 Earth masses for pebbles that are 50 percent ice by mass, increasing with opacity and pebble flux, and decreasing with pebble ice/rock ratio.

\end{abstract}


%
%
\section{Introduction} \label{sec:intro}

The formation of gas-giant planets like Jupiter is a central problem in planetary science. These objects contain large amounts of hydrogen and helium \citep{owen:1999}, which were presumably acquired from the gaseous portion of their star's protoplanetary disk. Most disks are found around stars less than a few million years old \citep{haisch:2001}, and this sets an upper limit on the time available to form gas giants. This time constraint is more severe than for rocky planets, such as Earth, which may acquire much of their bulk after the  protoplanetary disk disperses. 

Reproducing the existence and characteristics of gas giants is a challenging test for theories of planet formation. Moreover, understanding giant-planet formation is essential for understanding the origin and character of planetary systems in general, since gravitational perturbations from giant planets can effect the growth of objects elsewhere in the same system \citep{chambers:2002, levison:2003}. For example, perturbations from Jupiter and Saturn play a central role in many models for the origin of the Sun's asteroid belt \citep{wetherill:1992, chambers:2001, walsh:2011}. The gravitational influence of giant planets is further enhanced by the apparent tendency of many giants to migrate radially during or after their formation \citep{trilling:1999, armitage:2007}.

Usually, the formation of gas-giant planets is described using the ``core-accretion'' model. Here, a giant begins as a solid protoplanet that grows by accreting asteroid-sized planetesimals \citep{pollack:1996, movshovitz:2010}. Gas in the vicinity of the protoplanet is compressed and heated by the protoplanet's gravity, forming an extended atmosphere surrounding a solid core. The mass of the atmosphere is determined by a balance between the inward pull of gravity and the outward pressure gradient. Typically, the atmospheric mass increases rapidly with increasing core mass \citep{pollack:1996}. For a given set of conditions there is a critical core mass above which a static atmosphere is no longer possible \citep{mizuno:1980}. If the core exceeds the critical mass, gas flows into the atmosphere from the disk at an increasing rate, eventually forming a massive, gas-rich planet.

Multiple factors can affect the critical core mass, in particular the luminosity, opacity, and composition of the atmosphere \citep{stevenson:1982, hori:2011}. Lowering the opacity and/or luminosity increases the density in the atmosphere, raising the mass of the atmosphere for a given core mass. This in turn reduces the critical core mass. The main source of luminosity is energy released by infalling planetesimals, while opacity in the outer atmosphere is mostly due to dust embedded in the gas \citep{mizuno:1980}. Early studies of giant-planet formation typically assumed opacities appropriate for an interstellar dust size distribution, leading to large critical core masses, around 10--20 Earth masses \citep{pollack:1996}. More recent works that account for dust coagulation within the atmosphere find lower opacities and critical core masses below 10 Earth masses \citep{movshovitz:2010, mordasini:2014}.

Previous studies have mainly considered atmospheres with compositions similar to the Sun---that is roughly 99\% hydrogen and helium, with trace amounts of heavier elements \citep{pollack:1996, rafikov:2006, movshovitz:2010, piso:2014}. However, the atmospheric composition can be altered by the addition of material evaporating from planetesimals as they fall towards the core \citep{stevenson:1984, hori:2011}. The evaporated materials have higher molecular weights than hydrogen and helium, which can substantially increase the density of the atmosphere. For atmospheres dominated by heavy elements, the critical core mass can be smaller than Earth or even Mars \citep{hori:2011, venturini:2015}.

While conventional models consider cores that accrete planetesimals, some recent studies have suggested that planets acquire their solid material in the form of mm-to-m-sized pebbles instead \citep{lambrechts:2012, levison:2015, chambers:2016}.  There is some observational evidence  that pebbles are abundant in protoplanetary disks \citep{testi:2003, wilner:2005}. Pebble accretion may also alleviate the problem of forming critical-mass cores within the lifetime of a protoplanetary disk. Pebbles experience strong drag with disk gas during an encounter with a core, greatly increasing the capture probability and the core growth rate in some circumstances \citep{ormel:2010, lambrechts:2012}.

Outside a protoplanetary disk's ice line, pebbles are likely to be a mixture of volatile ices and refractory materials (silicates, metals etc., referred to here as ``rock''). Temperatures in a protoplanet's atmosphere increase with depth \citep{rafikov:2006}, so the icy component of incoming pebbles will evaporate in most cases, raising the molecular weight of the atmosphere and lowering the critical core mass. This process is likely to be much more effective than evaporation from planetesimals for  two reasons: (i) the surface area to volume ratio of pebbles is much greater due to the small size of pebbles, and (ii) pebbles typically settle towards the core relatively slowly, at terminal velocity, whereas planetesimals approach at roughly the escape velocity of the core. This  allows more time for evaporation to occur.

Relatively little evaporation will take place in the cool, outer regions of an atmosphere since the vapor pressure of water is low here, and the atmosphere soon becomes saturated. Above a certain temperature, however, the ice component of pebbles will evaporate entirely. Pebbles with a solar composition of condensible materials will have an ice-to-rock ratio of roughly unity \citep{lodders:2003}. Thus, cores that grow by accreting icy pebbles should have vapor-rich atmospheres that are more massive than the core itself, when the hydrogen/helium component is taken into account. This contrasts with the standard H/He dominated case where the critical mass usually occurs when the core and atmospheric masses are comparable \citep{pollack:1996}.

So far, the atmospheres of planets undergoing pebble accretion have received relatively little attention. In this paper, we investigate the growth and structure of the atmospheres of giant-planet cores below the critical mass as they accrete ice-rich pebbles. We examine the various stages that such planets pass through as they grow, and determine the point at which they reach the critical mass. This work extends previous studies in several ways: (i) using realistic estimates for the pebble accretion rate rather than adopting  a fixed luminosity, (ii) allowing for varying atmospheric composition with depth due to water condensation and precipitation, and (iii) including the presence of an ocean where appropriate. 

The rest of this paper is organized as follows. Section~2 describes the model used for the protoplanet's atmosphere and the rate of pebble accretion. Section~3 looks at the various evolutionary stages that a protoplanet and its atmosphere pass through as the mass increases. Section~4 examines the critical mass as a function of key model parameters. Section~5 contains a dicussion, and the main results are summarized in Section~6.

%
%
\section{The Model}
We consider a single protoplanet moving on a circular orbit in a gas-rich protoplanetary disk. A constant mass flux of pebbles drifts past the planet, moving towards the star due to gas drag. For simplicity, the pebbles are assumed to be composed of rock and water ice only. A fraction of the pebbles are captured by the planet. The rocky component of these pebbles is assumed to sediment to the planet's solid core. At each radial distance from the core, the pebbles' ice component is assumed to evaporate immediately until the local atmosphere is saturated in water vapor, or until the ice has evaporated completely. If the pebbles still contain some ice when they reach the base of the atmosphere this ice is added to the solid core, or forms an ocean entirely composed of liquid water directly above the core, depending on the temperature at this point.

Values for the most important parameters used in the nominal model are listed in Table~1. In particular, we consider a protoplanet orbiting at $a=3$ AU from a solar mass star. We assume this location lies just outside the ice line in the disk, so that the pebbles here contain a mixture of rock and water ice, while more volatile ices have evaporated from the pebbles. The local temperature and gas density are 160 K and $10^{-10}$ g/cm$^3$ respectively, and these values provide the outer boundary conditions for the protoplanet's atmosphere.

The planet's atmosphere is assumed to be in thermal and hydrostatic equilibrium. The temperature, pressure and density all increase inwards. The atmosphere is heated by the gravitational energy released by infalling pebbles, and  this is assumed  to be the only significant source of heating. In the inner regions of the atmosphere, this energy is typically transported outwards by convection, while heat transport in the outer atmosphere is by radiation. The outer regions of the atmosphere are assumed to be fully saturated with water vapor due to evaporation from incoming pebbles. For low-mass planets, the saturated atmosphere extends down to the surface of an ocean that contains the bulk of the water budget. When temperatures are too high for an ocean to exist, the outer atmosphere is still saturated with water vapor, but the inner regions may be undersaturated (if insufficient water is available), or above the critical temperature for water. 

Radial profiles for temperature $T$, pressure $P$, density $\rho$, and interior mass $M$ are calculated using the standard stellar structure equations:
\begin{eqnarray}
\frac{dM}{dr}&=&4\pi\rho r^2 \nonumber \\
\frac{dP}{dr}&=&-\frac{GM\rho}{r^2} \nonumber \\
\frac{dT}{dr}&=&-\frac{3\kappa L\rho}{64\pi\sigma_BT^3r^2}
\hspace{18mm} {\rm radiative}
\nonumber \\
\frac{dT}{dP}&=&\frac{\adb T}{P}
\hspace{30mm} {\rm convective}
\label{eq_main_equations}
\end{eqnarray}
\citep{inaba:2003}, where $r$ is the radius, $\kappa$ is the opacity, $L$ is the luminosity, $\sigma_B$ is the Stefan-Boltzmann constant, and $\adb$ is the adiabatic gradient. The luminosity is assumed to be independent of radius, and equal to the gravitational energy released by the rocky component of the pebbles falling to the solid core of the planet: 
\begin{equation}
L=\frac{GM_{\rm core}f_{\rm rock}}{R_{\rm core}}
\frac{dM}{dt}
\end{equation}
where $M_{\rm core}$ and $R_{\rm core}$ are the mass and radius of the solid part of the planet, $dM/dt$ is the pebble mass accretion rate, and $f_{\rm rock}$ is the rock mass fraction of the pebbles.

The atmosphere is assumed to be convective wherever the convective temperature gradient is shallower than the radiative gradient. In convective regions saturated with water vapor, the temperature profile is assumed to follow a moist adiabat due to condensation of water/ice. Condensed water (or ice) is assumed to immediately precipitate to lower, unsaturated regions of the atmosphere or the ocean\footnote{Since the temperature of a rising parcel of gas undergoing convection falls at the same rate as its surroundings, the saturation vapor pressures are the same in each case, and so are the water fractions. Thus we can use the Schwarzschild criterion for convection rather than the more general Ledoux criterion.}. Convective regions that are not saturated, or lie above the critical temperature of water, follow the usual dry adiabatic profile \citep{kasting:1988, venturini:2015}.

Following \citet{kasting:1988}, the moist adiabatic gradient is given by
\begin{equation}
\frac{d\ln P}{d\ln T}=\frac{1}{\adb}
=\frac{P_v}{P}\frac{d\ln P_v}{d\ln T}
+\frac{P_g}{P}\left[1+\frac{d\ln\rho_v}{d\ln T}-\frac{d\ln\alpha_v}{d\ln T}\right]
\end{equation}
where the subscripts $v$ and $g$ refer to water vapor and the non-condensible gas (hydrogen plus helium) respectively. Here, the derivatives of $P_v$ and $\rho_v$ are taken along the saturation vapor pressure-temperature curve of water, taken from \citet{haar:1984}. Also, $\alpha_v$ is the ratio of the water vapor density to the gas density, whose derivative is given by
\begin{equation}
\frac{d\ln\alpha_v}{d\ln T}=
\frac{k\rho_g (d\ln\rho_v/d\ln T)-\rho_g\mu_gm_Hc_g
-\rho_v\mu_gm_H(ds_v/d\ln T)}
{\rho_v\mu_g m_H\Delta s+k\rho_g}
\end{equation}
where $k$ is Boltzmann's constant, $m_H$ is the mass of a hydrogen atom, $\mu$ is the molecular weight, $c$ is the specific heat at constant volume, $s$ is specific entropy, and $\Delta s$ is the specific entropy difference between water vapor and condensed water (liquid or solid). Note that in deriving this equation, the non-condensible gas is assumed to be ideal in the saturated regions of the atmosphere. (See \citet{kasting:1988} for a detailed derivation of these equations.)

\startlongtable
\begin{deluxetable}{ll}
\tablecaption{Parameters used in the nominal model}
\tablehead{
\colhead{Parameter} & \colhead{Value} \\
}
\startdata
Stellar mass & 1 solar mass \\
Planet semi-major axis & 3 AU \\
Disk temperature & 160 K \\
Disk density & $10^{-10}$ g/cm$^3$ \\
Turbulent viscosity $\alpha$ & 0.001 \\
Rock density & 4 g/cm$^3$ \\
Ice/rock ratio & 1:1 \\
Hydrogen/helium ratio & 3:1 \\
Pebble flux & $10^{-5}M_\oplus$/y \\
Pebble Stokes number & 0.01 \\
\enddata
\end{deluxetable}

Following \cite{lissauer:2009}, the outer radius of the atmosphere is given by
\begin{equation}
r_{\rm max}=\min\left[\frac{r_H}{4}, r_B\right]
\end{equation}
where $r_H$ and $r_B$ are the Hill radius and Bondi radius, given by
\begin{eqnarray}
r_H&=&a\left(\frac{\mtot}{3M_\odot}\right)^{1/3}
\nonumber \\
r_B&=&\frac{G\mtot}{c_s^2}
\end{eqnarray}
where $\mtot$ is the total mass of the planet including its atmosphere, and $c_s$ is the local sound speed of the gas in the disk.

The opacity is given by the sum of contributions due to dust and gas, using relations developed by \cite{freedman:2014} for the gas opacity. We adopt a constant dust opacity that is a model parameter, using $\kappa_{\rm dust}=0.01$ cm$^2$/g in the nominal case. More detailed treatments that take into account dust coagulation in the atmosphere have been developed recently \citep{mordasini:2014}. However, these models do not include source and sink terms due to a large flux of pebbles passing through the atmosphere and evaporation/condensation of water, so we use the simpler, constant-$\kappa_{\rm dust}$ case here.

To solve the atmospheric structure equations, we also need an equation of state (EOS). Here we adopt EOS for hydrogen and helium developed by \citep{saumon:1995}. For temperatures below 2000 K, we use the water/steam EOS by \cite{haar:1984}. Above 3000 K, we use EOS data from \cite{french:2009}. In between these temperatures, we smoothly interpolate between these two cases. For water ice I, we follow \cite{feistel:2006}. High pressure phases of ices could also exist on some protoplanets, but we do not encounter these phases in the models presented here. The approximate density of a H/He/water vapor mixture is determined by calculating the density for each component separately using the appropriate EOS and the partial pressure of the component, then summing the densities.\footnote{A more commonly used approximation is $1/\rho=\sum_iX_i/\rho_i$ where $X_i$ is the mass fraction of component $i$, and $\rho_i$ is calculated using the total pressure rather than the partial pressure \citep{nettelmann:2008}. However this method is not feasible when one of the components will condense under the total pressure.}

We calculate the rate at which the protoplanet accretes pebbles using the relations developed by \cite{ormel:2010}, which depend on the Stokes number $\st$ of the pebbles, given by
\begin{equation}
\st=\Omega\tstop
\end{equation}
where $\Omega$ is the Keplerian orbital frequency and $\tstop$ is the stopping time due to gas drag.

Small pebbles settle towards the planet at terminal velocity, leading to a capture radius $r_c$ given by
\begin{equation}
\left(\frac{r_c}{r_H}\right)^3+\frac{2\vrel}{3\Omega r_H}\left(\frac{r_c}{r_H}\right)^2
-8\st=0
\end{equation}
where the average approach velocity $\vrel$ of the pebbles is given by
\begin{equation}
\vrel=\Omega\times\max\left[\eta a,\frac{3r_c}{2}\right]
\end{equation}
where $\eta\simeq(c_s/a\Omega)^2$ is the fractional orbital velocity difference between the gas and the protoplanet.

Pebbles with Stokes numbers larger than a critical value $\stcrit$ do not reach terminal velocity, but still experience an enhanced capture probability due to gas drag. \cite{ormel:2010} find empirically that the capture radius in this case is given by the expression above modified by a factor of $\exp[-(\st/\stcrit)^{0.65}]$, where
\begin{equation}
\stcrit=\min\left[1, 12\left(\frac{\Omega a}{\vrel}\right)^3\right]
\end{equation}

The growth rate of  the protoplanet is 
\begin{equation}
\frac{dM}{dt}=\min\left[2r_c,\frac{\pi r_c^2}{\hpeb}\right]\times\Sigma\vrel
\end{equation}

Here, $\hpeb$ is the scale height of the pebbles. Following \cite{youdin:2007}, we assume that $\hpeb$ is set by a balance between gravitational settling and turbulent stirring, and is given by
\begin{equation}
\hpeb=\hgas\left(\frac{\alpha}{\alpha+\st}\right)^{1/2}
\end{equation}
where $\hgas$ is the gas scale height, and $\alpha$ is the turbulent viscosity parameter \citep{shakura:1973}.

The surface density of pebbles $\Sigma$ is related to the pebble flux $F$ and the radial velocity $v_r$:
\begin{equation}
F=2\pi a\Sigma v_r
\end{equation}
where, following \cite{weidenschilling:1977}, we have
\begin{equation}
v_r=\frac{2\eta a\Omega\st}{1+\st^2}
\end{equation}

\begin{figure}
\centering
\includegraphics[height=120mm,angle=270]{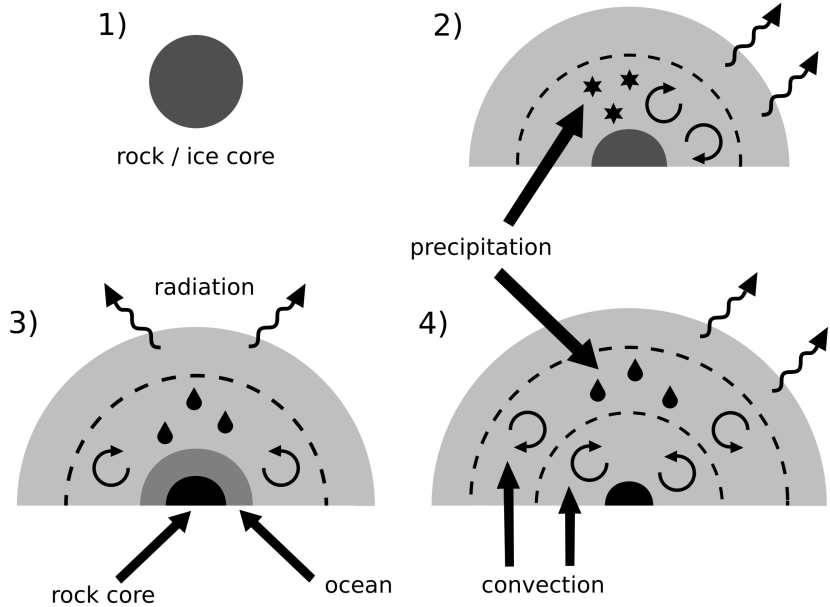}
\caption{Schematic diagram showing several evolutionary stages of a protoplanet accreting ice-rich pebbles. (1) Planet without an atmosphere. (2) Planet with an atmosphere and a solid ice/rock surface. The outer atmosphere is radiative. The inner atmosphere is convective with ice grains precipitating towards the surface. (3) Planet with a rocky core and an ocean. The outer atmosphere is radiative, and the inner atmosphere is convective with ice and water precipitation. (4) Planet with a rock core and no ocean. The outer atmosphere is radiative, and the inner atmosphere is convective. Precipitation occurs at mid altitudes, but the deep atmosphere is too hot for water to condense or undersaturated.}
\label{fig_schematic}
\end{figure}

%
%
\section{Nominal Case}
In this section, we consider the atmospheric structure of a growing protoplanet for the model parameters listed in Table~1. The planet and its atmosphere pass  through a series of evolutionary stages, some of which include an ocean. At each stage, we assume the body is in thermal and hydrostatic equilibrium. The main stages are illustrated schematically in Figure~1.

Initially, the protoplanet is too small to possess an atmosphere. Pebbles accreting onto the planet remain frozen until they settle to the surface, so that the planetary composition is the same as the pebbles themselves. The protoplanet first begins to acquire an atmosphere when its Bondi radius equals its physical radius. At this point, the planet's mass is
\begin{equation}
M=\frac{c_s^3}{G^{3/2}}\left(\frac{3}{4\rho_{\rm solid}}\right)^{1/2}
\end{equation}
For the nominal model parameters, and a planet composed of 50\% rock and 50\% ice, an atmosphere  first forms when the total mass reaches about $0.0017$ Earth masses. The atmosphere is extremely tenuous at first, but the density near the surface grows rapidly with increasing mass of the protoplanet.

When the total mass of the planet reaches about 0.084 Earth masses, the surface becomes warm enough for ice to melt. At this point an ocean begins to form. In practice, some time will be required to melt all of the planet's ice component, but here we assume an ocean forms instantaneously. The mass of H and He gas is very small at this stage, and the rocky core contains almost 50\% of the total mass, or 0.042 Earth masses.

\begin{figure}
\plotone{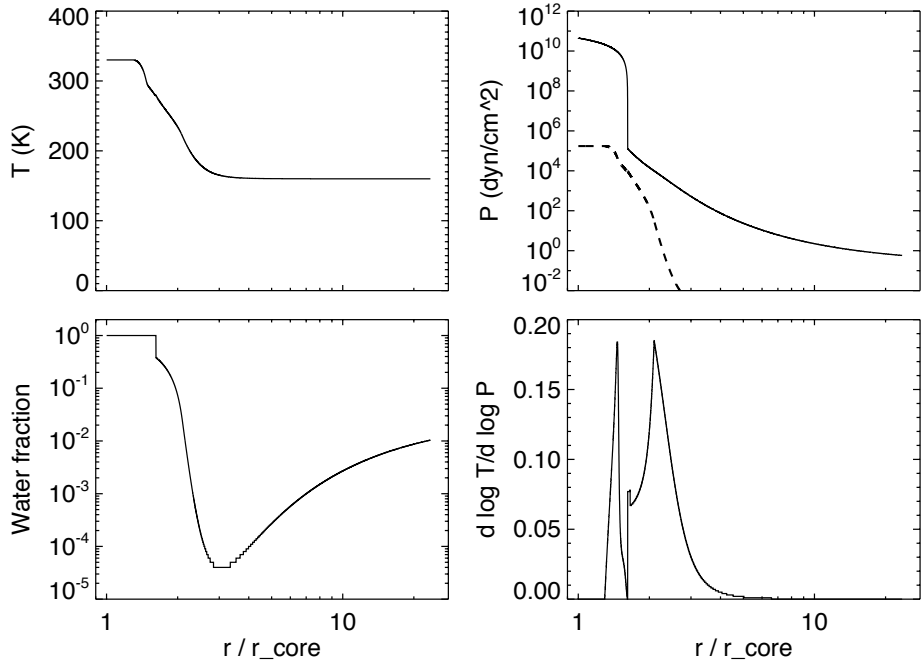}
\caption{The structure of the atmosphere and ocean for a rocky core with a mass of 0.042 Earth masses. The other model parameters are given in Table~1. The upper, left panel shows temperature. The upper, right panel shows total pressure (solid line), and the saturation vapor pressure of water (dashed line). The lower, left panel shows the water mass fraction. The lower, right panel shows the logarithmic temperature-pressure gradient.}
\label{fig_first_ocean}
\end{figure}

Figure~\ref{fig_first_ocean} shows the radial structure of the atmosphere and ocean in this case. Most of the atmosphere is nearly isothermal with a temperature similar to the surrounding protoplanetary disk. Energy in this region is transported efficiently by radiation since this part of the atmosphere is optically thin. Pressure increases roughly exponentially in the outer atmosphere. The low temperature means that the mass fraction of water is very low, even though the atmosphere is fully saturated with water vapor, as shown in the lower, left panel of the figure. The water mass fraction decreases to a minimum of about $4\times 10^{-5}$ at 3 core radii, before the rising temperature allows it to increase again at smaller radii.

At a radius of about 2.1 core radii, the atmosphere becomes convective. This results in a sharp change in slope of the temperature gradient with respect to pressure, shown in the lower, right panel of Figure~\ref{fig_first_ocean}. The atmosphere now enters a moist adiabatic regime in which gas rises and cools, causing water to condense and releasing latent heat. The release of latent heat reduces the adiabatic gradient as a result. Initially, the condensed water forms ice, which is assumed to precipitate immediately towards the core. Once the triple point is reached, excess vapor condenses as liquid water instead. This transition results in a small discontinuity in the temperature-pressure gradient due to the different entropy content of ice versus water.

At about 1.6 core radii, the atmosphere gives way to an ocean, which contains the bulk of the planet's water budget. The pressure and density jump by several orders of magnitude at this point, which can be seen in the upper, right panel of Figure~\ref{fig_first_ocean}. The ocean is convective since heat released by infalling pebbles is assumed to be deposited at the surface of the rocky core. However, the increase in temperature with ocean depth is small in this case.

As the mass of a protoplanet increases, the atmospheric pressure and temperature increase as well. A hotter atmosphere allows more water vapor to be present due to the higher saturation vapor pressure. The increase in temperature also means that the surface of the ocean grows hotter. Since the ocean itself is convective, the lower portions eventually become a super-critical fluid once the temperature at the surface of the core exceeds 647~K, the critical temperature of water. However, density and temperature vary smoothly across the critical transition.

\begin{figure}
\plotone{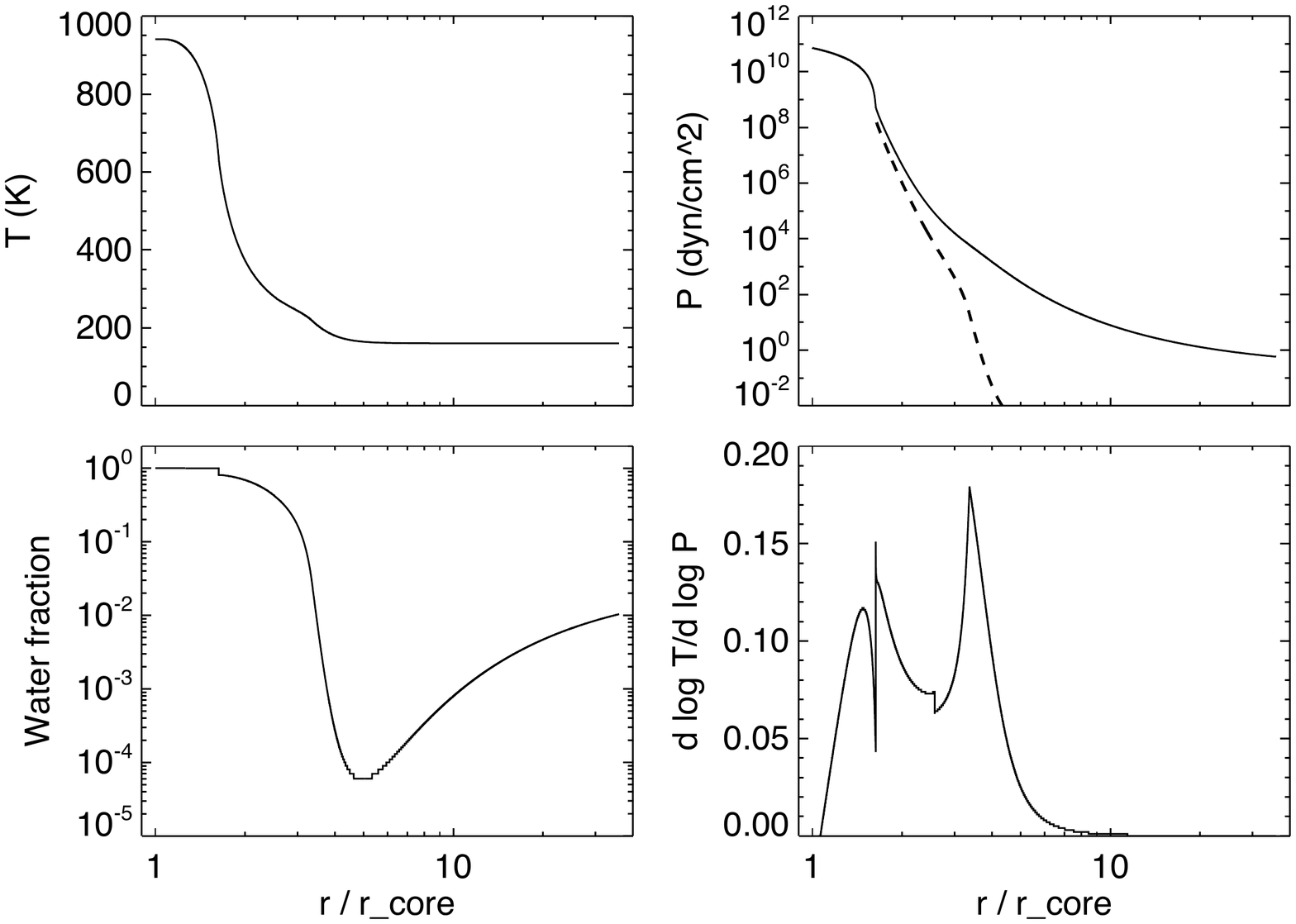}
\caption{The structure of the atmosphere and ocean for a rocky core with a mass of 0.0794 Earth masses. The other model parameters are given in Table~1.}
\label{fig_last_ocean}
\end{figure}

Figure~\ref{fig_last_ocean} shows the structure of the atmosphere and ocean for a protoplanet with a rocky core mass of 0.0794 Earth masses. For this core mass, the ocean surface temperature is 637~K, only slightly below the critical temperature. Thus, this core is one of the most massive that can sustain an ocean. The structure of the outer atmosphere in this case is similar to that in Figure~\ref{fig_first_ocean}. Heat is transported by radiation, the temperature is nearly constant, and the pressure increases exponentially with decreasing radius. As before, the low temperature means that the water mass fraction remains very low in this region.

The radiative-convective boundary in Figure~\ref{fig_last_ocean} occurs at about 3.4 core radii. This is larger than in Figure~\ref{fig_first_ocean} for two reasons. Firstly, the planet is more massive so the atmosphere is denser at a given distance from the core. This increases the temperature gradient in the radiative region (see the third of Eqns~\ref{eq_main_equations}). Secondly, the stronger gravity of the planet increases the efficiency of pebble accretion and the energy deposited by infalling pebbles. As a result, the luminosity is about 3 times higher than the case shown in Figure~\ref{fig_first_ocean}. The increased luminosity and atmospheric density both mean that convection must operate further from the core than before.

Interior to the radiative zone is a convective zone extending down to the ocean surface at 1.6 core radii. The convective atmosphere follows a moist adiabat, with the H/He gas saturated in water vapor. The temperature-pressure gradient decreases moving inwards until the temperature reaches the triple point of water, at 273~K, and then increases again. As before, the water mass fraction decreases with depth in the outer atmosphere, reaching a minimum of $6\times 10^{-5}$ at about 5 core radii. 

The higher temperatures in the inner atmosphere lead to much higher water fractions. By the time the ocean surface is reached, the atmosphere is 81\% water by mass compared to only 38\% in Figure~\ref{fig_first_ocean}. The large amount of water present at the base of the atmosphere, coupled with temperatures near the critical point, mean the discontinuity at the ocean surface is much less pronounced than the previous case. The density increases by only a factor of about 4 for example. The temperature and density continue to rise with depth in the convective ocean, passing seamlessly through the critical point, and reaching a temperature of 940~K and a density of 1.6  g/cm$^3$ at the rocky core.

At core masses slightly higher than in Figure~\ref{fig_last_ocean}, an ocean ceases to exist. Instead the entire water budget exists as either water vapor or a super-critical fluid. We assume that the water becomes intimately mixed with the hydrogen and helium throughout the resulting atmosphere. (In practice, some mixing may occur when the ocean is slightly below the critical temperature, but we do not attempt to model this effect here.) Below a particular transition temperature, the atmosphere is saturated in water vapor. Above this temperature, we assume the water mass fraction is constant, and set by the total amount of water available (which is equal to the mass of the rocky core in the nominal model). For rocky core masses below 0.105 Earth masses, this transition occurs at or close to the critical point, while for more massive cores, the transition occurs at progressively lower temperatures. Thus, more massive cores have lower maximum water fractions, and increasing fractions of H/He gas.

\begin{figure}
\plotone{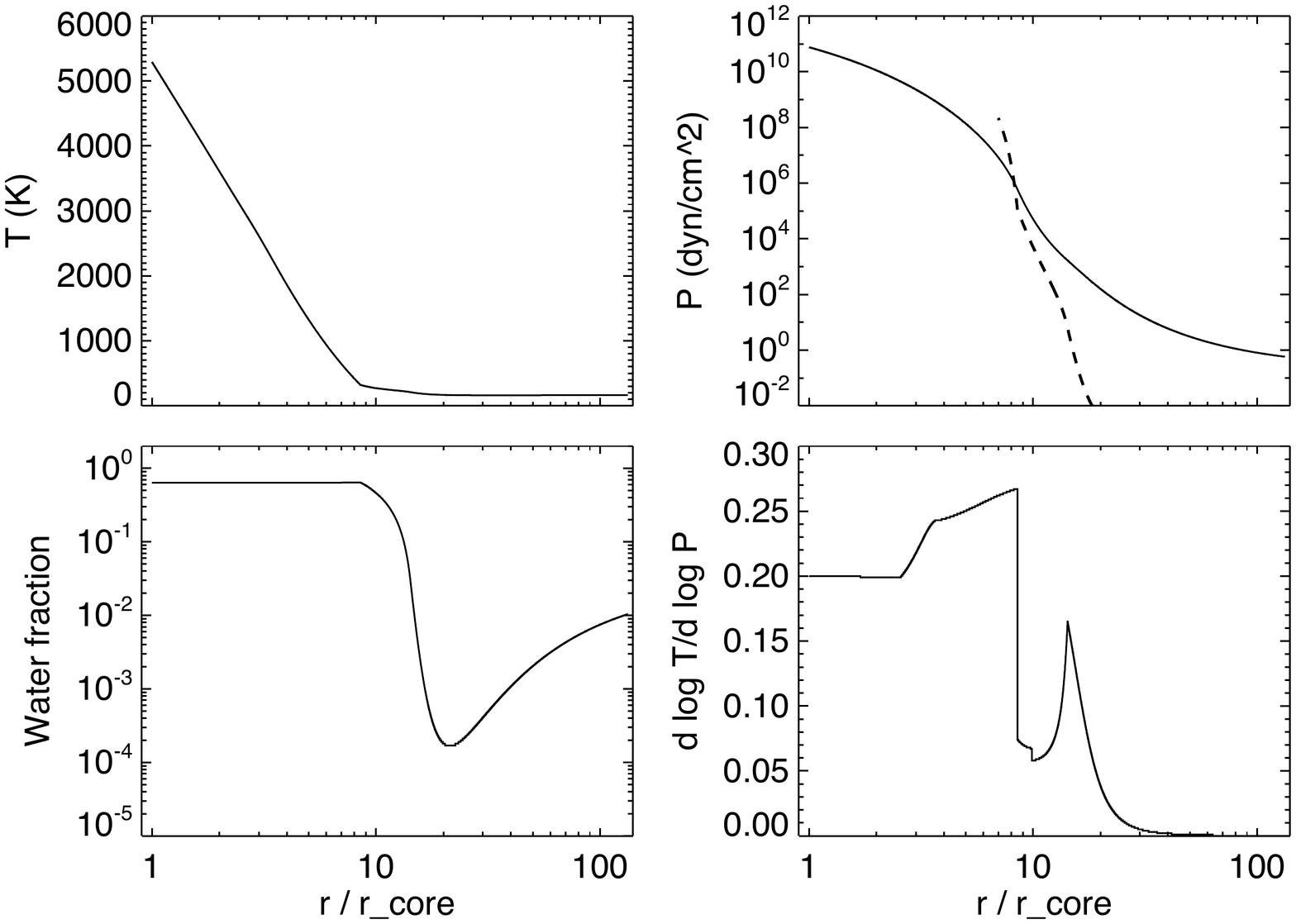}
\caption{The structure of the atmosphere and ocean for a rocky core with a mass of 0.391 Earth masses. The other model parameters are given in Table~1.}
\label{fig_no_ocean}
\end{figure}

Figure~\ref{fig_no_ocean} shows the atmospheric structure for a rocky core with a mass of 0.391 Earth masses. This object has a total mass (core plus atmosphere) of 1 Earth mass. The outer, nearly-isothermal part of the atmosphere looks similar to the previous cases. However, the radiative-convective boundary now lies at 14.3 core radii, much further out than before, due to the larger core mass and higher luminosity due to pebble  accretion. 

Inside the radiative-convective boundary lies a relatively narrow moist adiabatic region, extending inwards to 8.5 core radii, where the temperature is 321~K. Inside this point, the atmosphere becomes undersaturated in water vapor, as can be seen in the upper, right panel of Figure~\ref{fig_no_ocean}. At this radius, there is a large jump in the temperature-pressure gradient (see lower, right panel), and from this point inwards, the atmosphere follows a steeper, ``dry'' abiabat due to the absence of latent heat released by condensing water vapor.

Below 8.5 core radii, the water mass fraction is constant at 0.64, with the remaining mass made up of hydrogen and helium. The steep convective gradient means the temperature increases rapidly with depth, reaching 5900~K at the rocky core. (The changes in the temperature-pressure gradient seen at 2000~K and 3000~K in the lower, right panel of the figure are artifacts caused by changes in the EOS at these points.) The great majority of the atmospheric mass is contained in the region below 8.5 core radii, with only about 0.15\% of the total planetary mass lying at larger radii.

The trends described above continue for larger core masses, but eventually a point is reached where the core mass reaches a maximum. We examine this point in the next section.

\begin{figure}
\plotone{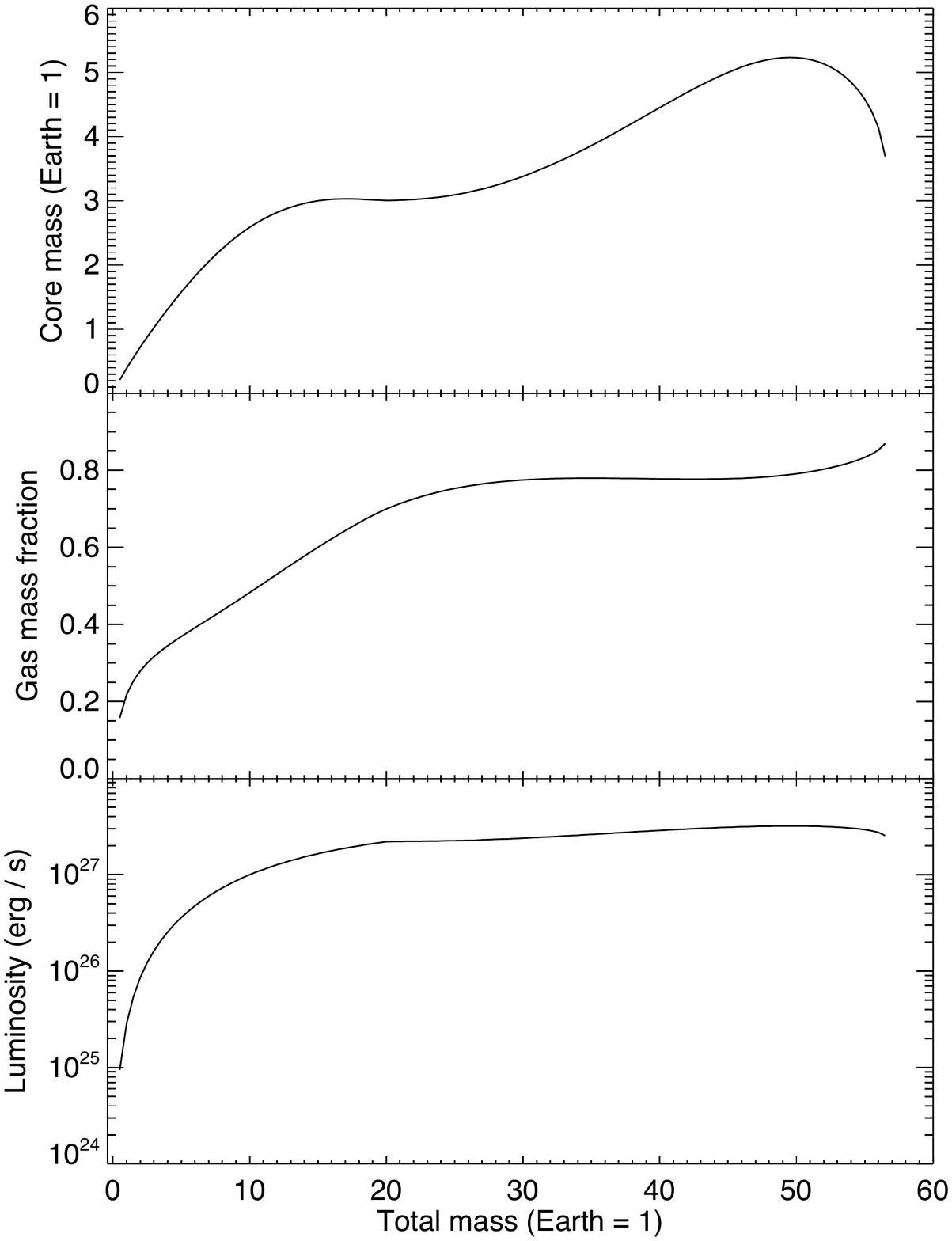}
\caption{Upper panel: rocky core mass as a function of total planet mass for the nominal model. Middle panel: gas mass fraction of these objects. Lower panel: luminosity due to pebble accretion.}
\label{fig_critical_mass}
\end{figure}

%
%
\section{Critical Core Mass}
Figure~\ref{fig_critical_mass} shows the change in atmospheric properties of protoplanets with increasing total mass for the parameters listed in Table~1. The top panel shows the rocky core mass as a function of the total planetary mass. The middle panel shows the gas (H plus He) mass fraction. The bottom panel shows the luminosity due to pebble accretion.

For total masses below about 10 Earth masses, the rocky core mass increases roughly linearly with total mass. A growing planet should follow this trajectory, with the total mass increasing smoothly with core mass. Above 10 Earth masses, the curve in the top panel of Figure~\ref{fig_critical_mass} starts to turn over. Small increases in core mass lead to progressively larger increases in atmospheric mass and thus total mass. At a total mass of 17 Earth masses, the core mass reaches a maximum value of 3.03 Earth masses. At this point H and He represent about 64\% of the total planetary mass.

Further increases in total mass result in a smaller core mass, before the core mass starts to increase again when the total mass reaches 20 Earth masses, eventually reaching a second maximum at a total mass of 50 Earth masses. In practice, the core mass should increase monotonically with time, so the region between 17 and 20 Earth masses is unphysical. It is possible that continued gas accretion will allow a real planet to move across this region before rejoining the solutions above 20 Earth masses. However, examining this possibility would require a model that includes time evolution and luminosity generated by gas accretion rather than the static model used here. 

It may seem surprising that a given core can support more than one type of atmosphere. However, the solutions differ in a number of respects. In particular, the luminosities are different because the capture radius for pebble accretion is determined by the total mass rather than the core mass. The atmospheric compositions are also different since a given mass of water is diluted to different degrees by the different masses of H/He gas for each atmospheric solution. We note that \cite{mizuno:1980} also found multiple atmospheric solutions for a given core mass. Here we follow \cite{mizuno:1980} and \cite{venturini:2015}, and associate the first core-mass maximum with the ``critical'' core mass, assuming that the static solutions become invalid beyond this point. 

A second criterion that is sometimes used to determine the onset of rapid gas accretion is the ``crossover'' mass \cite{pollack:1996, rafikov:2006}. This is the point at which the masses of the core and atmosphere are equal. In the model presented here, the atmospheric mass is always greater than the core mass, once an ocean ceases to exist, due to the presence of water vapor. Instead, we note the point at which H and He make up half of the total mass. Using this criterion, rapid gas accretion in Figure~\ref{fig_critical_mass} would begin at a core mass of 2.68 Earth masses, slightly below the critical mass.

In the following subsections, we will examine how the critical mass and crossover mass vary depending on some key model parameters.

%
%
\subsection{Dependence on Opacity}
One of the biggest uncertainties in the model is the opacity of dust grains in the atmosphere. Some dust is swept into the atmosphere with the inflowing gas. The opacity of this dust will depend on the local dust-to-gas ratio in the protoplanetary disk, which in turn depends on how much dust was converted into pebbles. A bigger uncertainty is the fate of infalling pebbles once their icy component evaporates. The rocky residues may remain intact, settling quickly towards the core and leaving the opacity unchanged. Alternatively, pebbles may disintegrate into sub-$\mu$m-size grains once the ice cementing these grains evaporates. In this case, the opacity could be raised by orders of magnitude. Intermediate scenarios are also possible. Given these uncertainties, we choose to treat the dust opacity as a parameter with a wide range of possible values.

\begin{figure}
\plotone{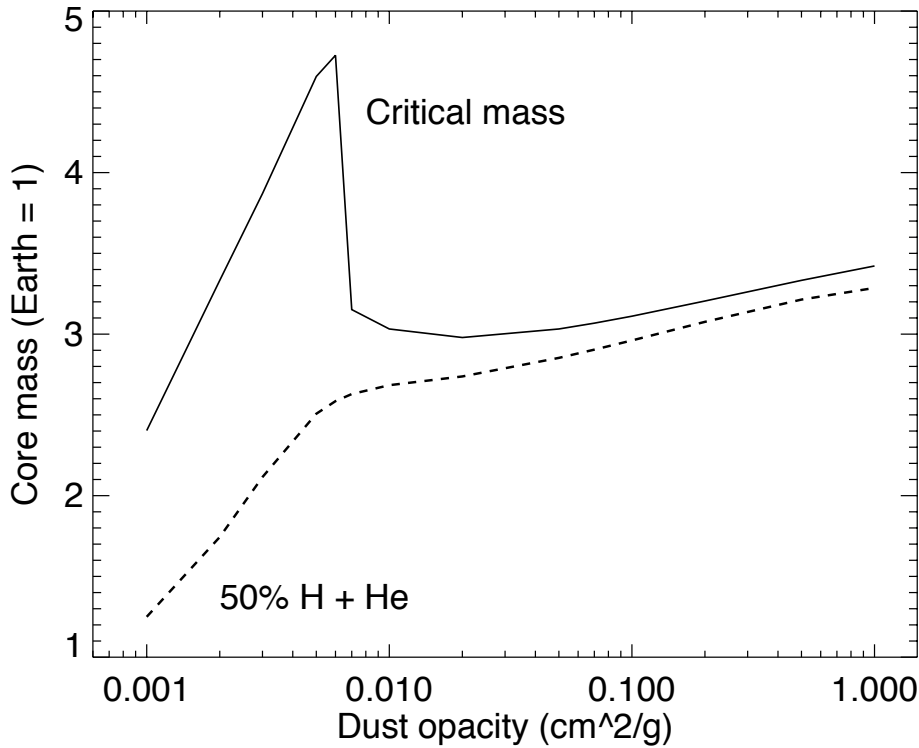}
\caption{Dependence of the critical core mass and crossover mass (50\% of the total mass in H and He) on the opacity of dust grains in the atmosphere.}
\label{fig_opacity}
\end{figure}

Figure~\ref{fig_opacity} shows the critical core mass and the crossover core mass as a function of the dust opacity $\kdust$. Other model parameters are unchanged from those in Table~1. The critical mass varies with $\kdust$ in a complicated manner. Initially, the critical mass increases with increasing $\kdust$ for small values, peaking at $\kdust=0.006$ cm$^2$/g, and then falling again. The critical mass becomes nearly constant at about 3 Earth masses for $\kdust>0.01$ cm$^2$/g. The crossover (50\% H and He) mass increases monotonically with dust opacity, rising rapidly for $\kdust<0.007$ cm$^2$/g, and more slowly at larger opacities.

The complicated behavior seen in the critical core mass is partly due to the occasional existence of multiple solutions for the core mass as a function of total mass, as seen in Figure~\ref{fig_critical_mass}. For most values of $\kdust$ in Figure~\ref{fig_opacity}, a single maximum exists, and this is the critical core mass. However, for intermediate values of $\kdust$ a second maximum appears at smaller core masses than the first. In these cases we assign the first maximum to be the critical mass, leading the decrease in critical mass seen at intermediate values of $\kdust$ in Figure~\ref{fig_opacity}. 

The crossover mass doesn't exhibit the complicated behavior shown by the critical mass, (The change of slope seen in the figure coincides with a change in the outer boundary condition from the Bondi radius to 1/4 of the Hill radius.) For this reason, the crossover mass may represent a more robust measure of the onset of rapid gas accretion than the critical mass. It is worth noting that the critical and crossover masses are similar for large dust opacities.

Overall, the core mass at the onset of rapid gas accretion tends to increase with increasing dust opacity. A higher opacity is associated with a steeper rise in temperature in the radiative region of the atmosphere. As a result, the boundary between the radiative and convective  regions lies further from the core when the opacity is large. In the radiative region, pressure and density both increase roughly exponentially moving inwards, while these quantities vary more slowly in the convective part of the atmosphere (see Figure~\ref{fig_no_ocean}). Since the total atmospheric mass depends on the integrated density, a smaller radiative region typically leads to a lower total mass for a given core mass, increasing the core mass needed for rapid gas accretion to begin.

\begin{figure}
\plotone{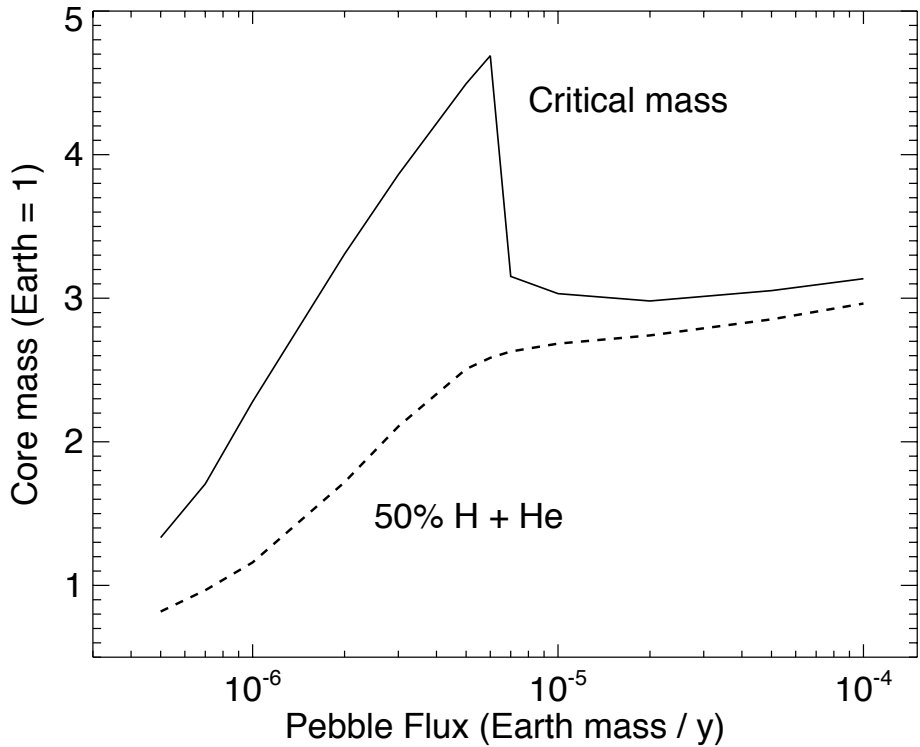}
\caption{Dependence of the critical core mass and crossover mass (50\% of the total mass in H and He) on the radial mass flux of pebbles drifting through the disk.}
\label{fig_pebble_flux}
\end{figure}

%
%
\subsection{Dependence on Pebble Flux}
The mass flux of pebbles passing  through the region containing the planet is another source of uncertainty. Pebble drift rates depend on the size of the pebbles and the radial pressure gradient in the disk. The pebble flux depends on the drift rate, and also on the total mass of pebbles in the disk, how long they have been drifting, and the radial extent of  the disk. The pebble flux is likely to vary with time and location, possibly by large amounts. In our model, the luminosity of the planet's atmosphere is determined by the pebble accretion rate, which depends on the pebble flux as well as the capture radius. Thus, we consider a range of possible pebble fluxes.

Figure~\ref{fig_pebble_flux} shows the critical core mass and crossover mass for a range of pebble mass flux values. The dependence of both masses is broadly similar to the dependence on the dust opacity shown in Figure~\ref{fig_opacity}. The critical mass increases with increasing flux, reaches a peak at intermediate values, then declines to a roughly constant value for large mass fluxes. The crossover mass increases monotonically with increasing flux, rapidly at first and  then slowly. For large pebble fluxes the two masses are similar.

The similar behavior in Figures~\ref{fig_opacity} and \ref{fig_pebble_flux} occurs because the opacity and luminosity (which is proportional to the pebble flux) have the same effect on the atmosphere. The temperature profile depends linearly on each quantity in the radiative region, while neither quantity directly affects the profile in the convective region, as can be seen in Eqns.~\ref{eq_main_equations}. The main difference between dust opacity and pebble mass flux is that the rate at which the planetary core accretes mass cannot exceed  the pebble flux itself, and this will be a factor in the next subsection.

\begin{figure}
\plotone{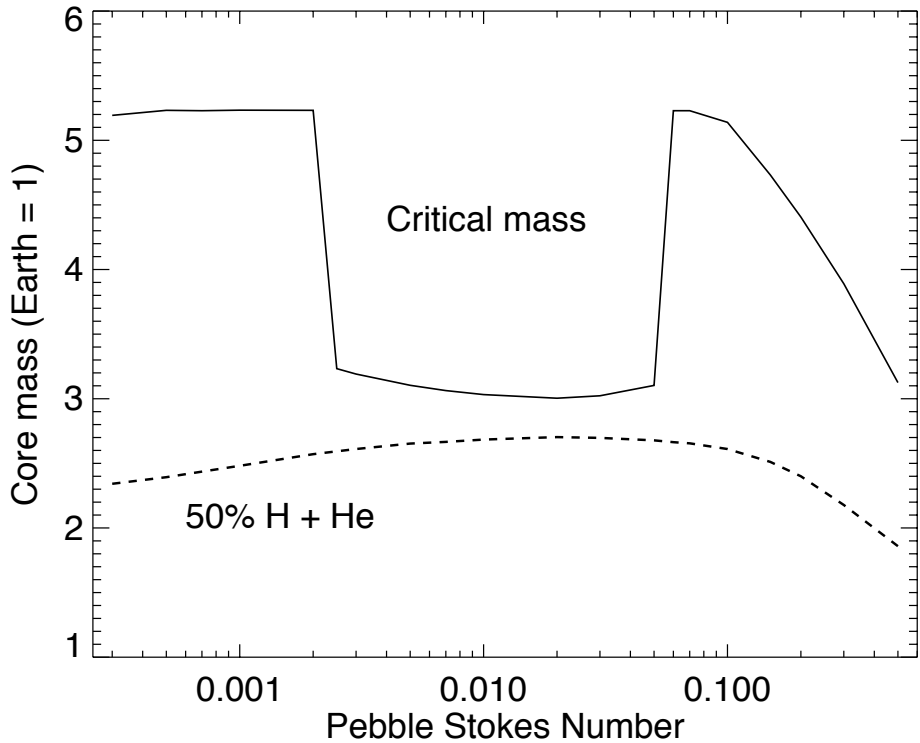}
\caption{Dependence of the critical core mass and crossover mass (50\% of the total mass in H and He) on the Stokes number of the pebbles.}
\label{fig_pebble_stokes}
\end{figure}

%
%
\subsection{Dependence on Pebble Size}
The size of pebbles in the planet-forming regions of protoplanetary disks is also uncertain, and probably varies with time and location. The typical Stokes number of pebbles will also vary depending on the local gas density. Here we examine a range of possible Stokes numbers $\st$. 

Figure~\ref{fig_pebble_stokes} shows the critical core mass and crossover mass as a function of $\st$. For $\st<0.002$, the critical mass is 5.2 Earth masses, almost independent of the Stokes number. For $0.002<\st<0.06$, a second maximum appears in the relation between core mass and total mass. For this range of Stokes numbers, the critical mass falls to roughly 3 Earth masses. At larger Stokes numbers, the critical core mass returns to the previous value, before declining with $\st$ for $\st>0.1$.

The constant critical mass for $\st<0.002$ arises because in these cases the pebble capture radius is large enough that the entire flux of pebbles is captured by the planet. The radial drift speeds of the pebbles are slow enough that they are all accreted by the planet in the time it takes them to drift inwards across the planet's Hill radius. As a result, changing the pebble Stokes number has no effect on the luminosity of the atmosphere or the atmospheric structure. 

For larger pebbles (larger Stokes numbers), the behavior of the critical mass is more complicated due to the appearance of an additional core-mass maximum as was the case in the previous sections. Here, the accretion rate is less than the pebble flux, so differences become apparent for different values of $\st$. As before, the crossover mass behaves in a simpler manner than the critical mass. The crossover mass depends only weakly on $\st$. Although the pebble capture radius tends to increase with increasing pebble size, this effect is offset by the greater radial drift speeds, which lowers the pebble  surface density by a comparable amount. As a result, the onset of rapid gas accretion probably depends only weakly on pebble size for a given mass flux of pebbles.

\begin{figure}
\plotone{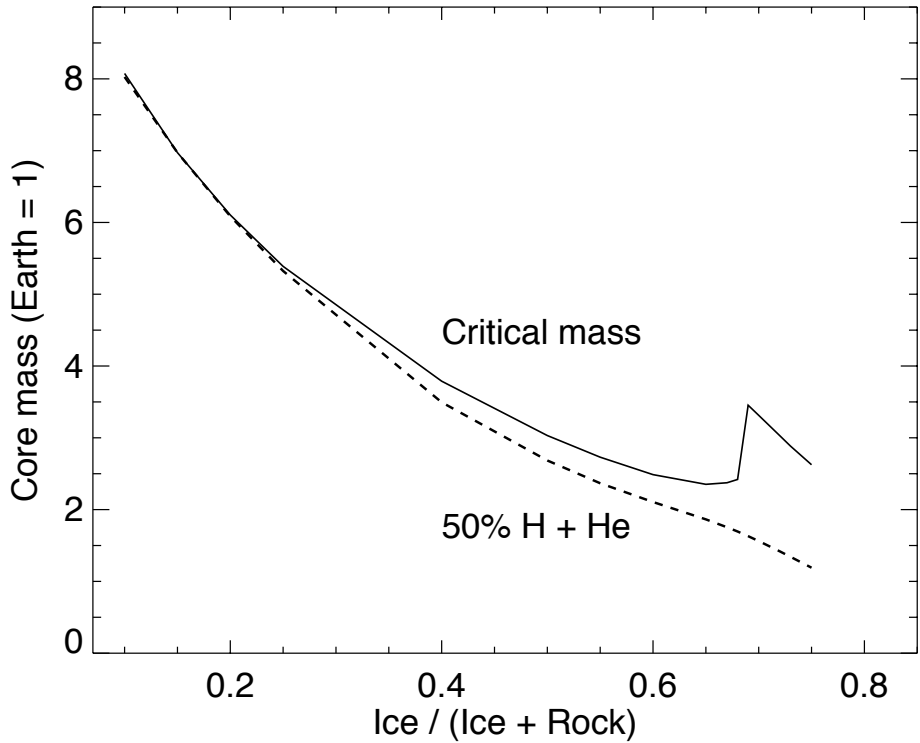}
\caption{Dependence of the critical core mass and crossover mass (50\% of the total mass in H and He) on the mass fraction of ice in the pebbles.}
\label{fig_ice_fraction}
\end{figure}

%
%
\subsection{Dependence on Ice-to-Rock Ratio}
So far, we have considered changes that alter  the structure of a planetary atmosphere's outer, radiative zone via the opacity or luminosity. In this section, we examine differences caused by altering the ice-to-rock ratio of the pebbles accreted  by the planet, which tends to affect the lower atmosphere. In the previous cases, we assumed that the ice mass fraction is 50\%. However, other values are possible for the solar nebula and other protoplanetary disks. In the solar nebula for example, the ice-to-rock ratio is somewhat uncertain due to uncertainties in the carbon-to-oxygen ratio of the Sun, and uncertainties regarding the main carrier of carbon in the solar nebula \citep{asplund:2009, lodders:2003}.

Figure~\ref{fig_ice_fraction} shows the critical core mass and crossover mass as a function of the ice mass fraction of pebbles. Unlike the other model parameters, both the critical mass and crossover mass show a clear, nearly monotonic trend with ice fraction. The masses decline steadily with ice fraction for fractions less than about 0.65, falling from 8.1 Earth masses to about 2.4 Earth masses for the critical mass and 1.9 Earth masses for the crossover mass. At this point, the critical mass increases somewhat, before resuming the previous trend. The crossover mass declines monotonically with ice fraction for all values considered here.

The strong dependence of the critical mass on the ice fraction comes from the fact that the atmospheric mass depends on the mean molecular weight, and this is greater when large amounts of water are present. When the mean molecular weight is high, the atmospheric mass is larger for a given core mass, leading to a smaller  critical mass. This is true only in the hot inner regions however. The effect seen in Figure~\ref{fig_ice_fraction} would be even stronger if the amount of water in the outer atmosphere wasn't restricted by the saturation pressure of water.

%
%
\section{Discussion}
In this paper we have examined the growth and atmospheric structure of planets accreting ice-rich pebbles in a protoplanetary disk of H and He gas. Such planets pass through several evolutionary stages, as shown in Figure~\ref{fig_schematic}, beginning with objects too small to appreciably affect the gas in the surrounding disk. When the planet's mass reaches about 0.002 Earth masses, it begins to acquire an atmosphere. The atmospheric mass increases rapidly with increasing core mass. The outer regions are optically thin and nearly isothermal, while close to the core temperatures increase rapidly and a convective zone develops.

Incoming pebbles sediment slowly towards the core at terminal velocity. As they encounter warmer regions, the icy component of the pebbles begins to evaporate, saturating the atmosphere with water vapor. Excess water and ice precipitate towards the core, forming an ocean when the surface temperature and pressure exceed the triple point of water. For the model parameters used here, an ocean first forms when the mass of rock reaches about 0.04 Earth masses. In the convective region, water vapor in the saturated gas condenses as the gas rises and cools. The condensed water precipitates back to the ocean, while the latent heat released by condensation means that the rising gas follows a shallow, moist-adiabatic temperature profile.

When the rocky core mass reaches about 0.08 Earth masses, the temperature becomes too high for an ocean to exist. We assume that water in the region closest to the rocky core forms a super-critical convecting fluid that mixes with gas in the layers above. Temperatures decline with distance from the core, passing smoothly through the critical point of water. The outer part of the convective zone continues to follow a moist adiabat. However, the higher temperatures in the inner part of the convective zone mean that the gas is undersaturated in water vapor, and here the temperature profile follows a steeper dry adiabat. It is possible that water from the ocean doesn't fully mix with the gas in the layers above when the ocean boils due to the steep molecular weight gradient. However, this probably won't make much difference to the outcome since the planetary mass at this stage is much lower than the critical mass. The amount of water that remains trapped in such a ``core'' is much less than the water supplied by more infalling pebbles, and this additional water will be mixed with gas in the convecting region of the atmophere.

As the core mass increases beyond 0.08 Earth masses in our model, the atmospheric mass continues to grow, and the mass ratio of water vapor to gas decreases. The region of the atmosphere that is saturated and convecting becomes increasingly narrow as a result. At a core mass of a few Earth masses, 50\% of the total planetary mass is made up of H and He gas. At a slightly higher mass, the rocky core mass reaches a maximum, and we assume the static model used here ceases to be valid at this point. Presumably, rapid gas accretion will begin to take place at around this critical mass, eventually forming a gas-giant planet.

The critical mass often exhibits complicated behavior as a function of the main model parameters. This is due to the existence of multiple maxima in the relation between total mass and core mass. A more robust indicator of the onset rapid gas accretion may be the crossover mass (defined here as the point when 50\% of the planet's mass is H and He), since this typically varies in a simpler manner when the parameters are varied. Better estimates of the onset of rapid gas accretion will come from evolving models that include the energy released due to gas accretion.

\cite{hori:2011} and \cite{venturini:2015} have carried out similar studies to the one in this paper, investigating the effects of evaporating volatiles on the atmospheric structure and critical mass of protoplanets. \cite{hori:2011} considered a two-layer atmosphere: an outer region with composition similar to the protoplanetary disk, and an inner region uniformly enriched in elements heavier than He. These authors assumed all volatile constituents remained in the vapor phase. \cite{venturini:2015} considered a single, uniformly enriched layer, and accounted for condensation of ice and water, but not precipitation.

\cite{hori:2011} found that the critical core mass typically decreases with increasing heavy-element concentration in the atmosphere for heavy-element mass fractions $Z>0.1$. For mass fractions above about 0.75, the critical mass is less than one Earth mass, and approaches Mars's mass for atmospheres almost wholly composed of heavy elements. \cite{venturini:2015} find similar results in the absence of water condensation. When condensation is included, they find that the critical mass is reduced further, falling below 0.1 Earth masses when $Z>0.75$.

In this paper, we find larger critical core masses, typically a few Earth masses, even though large amounts of water are present due to the evaporation of ice from pebbles. There are several reasons for this difference. Firstly, we consider only a single volatile, water, whereas \cite{hori:2011} and \cite{venturini:2015} included carbon-bearing species. The presence of carbon compounds, and transitions between the dominant species with changing temperature, lead to modest differences in atmospheric structure due to differences in the adiabatic temperature gradient. In addition, \cite{hori:2011} and \cite{venturini:2015} consider a fixed atmospheric luminosity $L=10^{27}$ erg/s, independent of planetary mass. Here, we calculate the luminosity self-consistently based on the pebble accretion rate, typically yielding slightly higher values at the critical mass.

However, the main reason for the larger critical masses found here is that we consider atmospheres with water mass fractions that vary considerably with altitude due to precipitation. Temperatures in the outer, radiative region are only slightly higher than in the local protoplanetary disk. As a result, the saturation pressure and mass fraction of water vapor are very low in this region. \cite{venturini:2015} assumed that condensed ice and water would remain where it forms, presumably in the form of clouds. We believe this is unlikely, particularly for the large water mass fractions they considered, and  that most of the condensed water will precipitate to the ocean or undersaturated regions deeper in the atmosphere. In fact, if mixing in the radiative region is inefficient, the water abundance may always be small since incoming pebbles will not evaporate where the gas is saturated in water vapor.

The very low abundance of water vapor in the upper atmosphere means that the radiative region and the upper convective region behave somewhat like the corresponding regions in conventional models for core accretion that consider a solar mixture of gases dominated by H and He, and yield large critical core masses \citep{pollack:1996, movshovitz:2010}. The presence of higher water fractions deep in the atmosphere clearly reduces the critical core mass, as can be seen in Figure~9, but the effect is more modest than found by \cite{hori:2011} and \cite{venturini:2015}. One difference between the conventional models and the one described here is that we assume that convection in the outer atmosphere begins when the temperature gradient reaches a moist adiabat, which is typically shallower than the dry adiabat that is usually considered. As a result, the radiative-convective boundary lies at a higher altitude in the model described here, raising the critical mass somewhat. Overall, however, previous estimates of the critical mass based on H- and He-dominated atmospheres may still be reasonably valid when the effect of ice evaporation from pebbles is taken into account.

In our model, two sources of negative feedback act to raise the critical mass even as the core itself grows. Firstly, an increase in the planetary mass increases the luminosity of the atmosphere, both because the capture cross section for pebbles becomes larger, and because the gravitational energy released by each infalling pebble becomes greater as the core's potential well deepens. Secondly, as the planet grows, the water-to-gas ratio in the atmosphere declines. Thus, the density effect due  to water vapor is increasingly diluted, and the presence of water vapor becomes less important for the structure of the atmosphere. Both factors need to be taken into account when determining the onset of rapid gas accretion.

In this study, we have considered only a single volatile species heavier than He, namely water. The rationale for this choice is that we consider a planet located just beyond the ice line that is accreting pebbles. While large planetesimals in this region may retain other, more-volatile ices, these ices will surely have evaporated from pebble-size particles before they encounter the planet. One possibility not considered here is that protoplanetary disks may contain an appreciable mass of tar-like organic compounds \citep{lodders:2004}, which are intermediate in volatility between water ice and rock. Such tars would be present in pebbles just outside the ice line, but would evaporate in the hotter regions of the planet's atmosphere. The presence of large amounts of tar-like material could alter the atmospheric structure and introduce some of the effects found by \cite{hori:2011} and \cite{venturini:2015}. Furthermore, the high temperatures we find close to the core of the planet may be sufficient for rocky materials to evaporate and decompose rather than sedimenting to the core as we have assumed here. These possibilities are worth studying further in future.

Some other factors may warrant further study. For example, the current model considers a constant mass flux of pebbles passing through the protoplanetary disk, and uses a fixed Stokes number for these pebbles. Neither quantity is likely to be constant in real disks \citep{birnstiel:2012}. The pebble flux probably declines over time, and the typical pebble size is likely to decrease as big pebbles are preferentially lost due to radial drift. Both factors will decrease the pebble capture rate and luminosity of the planet over time. Another factor we have not considered here is that the supply of pebbles may be shut off entirely once the planet becomes massive enough to alter the sign of the radial pressure gradient in nearby regions of the disk \citep{lambrechts:2014}.

In addition, we have assumed that the temperature and density at the outer boundary of the planet remain fixed. These quantities could change over time as the disk evolves or the planet migrates through the disk. Previous works have noted that the critical core mass is insensitive to the outer boundary conditions provided that the outer atmosphere is radiative \citep{stevenson:1982, rafikov:2006}. However, this will no longer be true if the outer boundary crosses the condensation front for a major species such as the ice line, since the composition of infalling pebbles will change abruptly at this point. In particular, a planet moving inwards across the ice line may change from a situation in which there is a net gain of water to a net loss as the water fraction in the outer atmosphere increases above the local disk value.

%
%
\section{Summary}
In this paper, we have modeled  the structure and critical mass of a planetary  atmosphere when the planet is accreting ice-rich pebbles. The main findings are:
\begin{enumerate}
\item The planet goes through a series of stages with increasing core mass. An atmosphere first appears for a core mass of 0.002 Earth masses. For core masses (rock + ice) between 0.08 and 0.16 Earth masses, the surface temperature is between the triple point and critical point of water, and an ocean forms. At larger masses, temperatures are too high for an ocean to exist, and water exists as a super-critical fluid mixed with H and He gas accreted from the protoplanetary disk.

\item Temperature in the atmosphere increases with depth. Some or all of the ice component of pebbles evaporates in the planet's atmosphere adding water vapor to the H and He gas from the disk. The saturation pressure of water is very low in the outer atmosphere, so the water mass fraction in this region is small. The water mass fraction is higher in the deep atmosphere, becoming constant below a certain altitude.

\item The outer atmosphere is radiative, the inner atmosphere is convective. In saturated, convective regions, the temperature profile follows a shallow, moist adiabat due to water condensation in the rising gas. Unsaturated regions follow a steeper dry adiabat. As the core mass increases, the water-to-gas fraction in the atmosphere declines, and boundary between unsaturated and saturated regions moves to lower temperatures and higher altitudes.

\item The mass of the planet's rocky core increases with total planetary mass. For a 1:1 ice-to-rock ratio, the core mass reaches a maximum value (the critical mass) at 2--5 Earth masses, and the static model used here is no longer valid. Presumably, runaway gas accretion begins at this point.

\item The critical mass depends in a complicated manner on the pebble size, mass flux through the disk, and atmospheric dust opacity. This is due to the presence of more than one core-mass maximum for a given total planetary mass for some parameter values. The critical mass declines nearly monotonically with increasing ice-to-rock ratio from 8 Earth masses at an ice fraction of 0.1 to about 2.5 Earth masses at an ice fraction of 0.75.

\item The point at which 50\% of the planetary mass is H and He (the crossover mass) varies with model parameters in a more straightforward way than the critical mass, and may be a better indicator of the onset of rapid gas accretion. The crossover mass increases with atmospheric dust opacity and pebble mass flux, and decreases with ice-to-rock ratio. The crossover mass is relatively insensitive to the pebble size.

\end{enumerate}

%
%
\acknowledgments

I would like to thank Alan Boss and Lindsey Chambers for helpful comments and discussions during the preparation of this paper. I also thank an anonymous referee whose comments have improved this paper.


%
%

\end{document}